\documentclass[runningheads]{llncs}
\usepackage{graphicx}
\usepackage{multirow}
\usepackage[symbol]{footmisc}
\usepackage[pdftex]{hyperref}

\begin{document}
\title{Automatic Brain Tumour Segmentation and Biophysics-Guided Survival Prediction}
\titlerunning{Brain Tumour Segmentation and Survival Prediction}

\author{Shuo Wang \inst{1,\dagger} \and
Chengliang Dai \inst{1,\dagger, \ddagger} \and
Yuanhan Mo \inst{1} \and \\ 
Elsa Angelini \inst{2} \and
Yike Guo \inst{1} \and
Wenjia Bai \inst{1,3}}
\footnotetext[2]{The two authors contributed equally to this paper.}
\footnotetext[3]{\email{c.dai@imperial.ac.uk}}
\authorrunning{S. Wang and C. Dai et al.}

\institute{Data Science Institute, Imperial College London, London, UK
\and
ITMAT Data Science Group, Imperial College London, London, UK
\and
Department of Brain Sciences, Imperial College London, London, UK
}

\maketitle

\begin{abstract}
Gliomas are the most common malignant brain tumours with intrinsic heterogeneity. Accurate segmentation of gliomas and their sub-regions on multi-parametric magnetic resonance images (mpMRI) is of great clinical importance, which defines tumour size, shape and appearance and provides abundant information for preoperative diagnosis, treatment planning and survival prediction. Recent developments on deep learning have significantly improved the performance of automated medical image segmentation. In this paper, we compare several state-of-the-art convolutional neural network models for brain tumour image segmentation. Based on the ensembled segmentation, we present a biophysics-guided prognostic model for patient overall survival prediction which outperforms a data-driven radiomics approach. Our method won the second place of the MICCAI 2019 BraTS Challenge for the overall survival prediction.
% \keywords{Brain imaging \and Deep learning  \and Tumour segmentation \and Radiomics}
\end{abstract}

\section{Introduction}
Gliomas are the most common malignant brain tumours in adults, characterised by intrinsic heterogeneity and dismal prognostics \cite{ricard2012primary}. Sub-regions with various biological properties coexist within the tumour and cause inconsistent treatment response. Multi-parametric magnetic resonance imaging (mpMRI) provides valuable information for characterising gliomas and their sub-regions, such as necrosis (NCR), non-enhancing tumour (NET), enhancing tumour (ET) and peritumoural edema (ED). Imaging phenotypes of these sub-regions show great potential in patient stratification \cite{li2019multi}. However, due to the highly heterogeneous shape and appearance, accurate segmentation of the tumour sub-regions requires expertise from experienced radiologists.

Automatic segmentation of brain tumour has drawn a lot of attention in the recent years due to the availability of open medical image datasets and the rapid development of convolutional neural networks (CNNs). A well-trained CNN model can finish the segmentation task in minutes with acceptable accuracy. However, there are still a few challenges including limited manually-annotated training data, variations in image acquisition protocols and MRI scanners etc. Apart from challenges for image segmentation, another challenge lies in building a robust prognostic model from the high-dimensional medical image phenotypes. Data-driven radiomics approach has demonstrated promising results while the reproducibility and explainability are still questionable \cite{scialpi2019radiomic}. 

To push the boundaries of automatic segmentation and survival prediction, Multimodal Brain Tumour Segmentation Challenge (BraTS) has been organised for the recent few years \cite{bakas2018identifying,bakas2017segmentation,bakas2017segmentation2,bakas2017advancing,menze2014multimodal}. The BraTS datasets consist of mpMRI scans for glioblastoma (GBM/HGG) and low grade glioma (LGG). The modalities include T1-weighted scan (T1), post-contrast T1-weighted scan (T1Gd), T2-weighted scan (T2) and T2 Fluid Attenuated Inversion Recovery (T2-FLAIR) scan. These scans were acquired preoperatively with different clinical protocols from multiple institutions and annotated by experienced radiologists.

In this paper, we compare several different neural network models for brain tumour image segmentation on the BraTS 2019 dataset and investigate the influence of attention units, loss function and post-processing on segmentation performance. Based on the ensembled segmentation, we develop a robust biophysics-guided model for survival prediction.

\section{Methods}
In this section, we first present our segmentation models. Based on the tumour sub-region segmentation, we define a series of tumour features, perform feature selection and finally propose a prognostic model.

\subsection{Tumour sub-region segmentation}
\subsubsection{Background: Winning methods in BraTS 2017 and 2018}
UNet and UNet-like models are adopted by most of the top participants in BraTS 2017 and 2018. \cite{myronenko20183d} won the first place in BraTS 2018 with a UNet architecture plus an additional decoder branch derived from a generic variational autoencoder (VAE). Given this modified network structure, the loss function used by \cite{myronenko20183d} consists of a soft Dice loss for the segmenter branch, the KL divergence loss and reconstruction loss for the reconstruction branch. A patch size of 160x192x128 was chosen to make the most use of the Nvidia Tesla V100 graphic card with 32GB GPU memory. The second place of BraTS 2018 used a vanilla UNet with minor modifications tailored for the BraTS dataset. The network was trained with both the BraTS dataset and an auxiliary dataset from their own institution to improve the Dice score of enhanced tumour region.

The winner of BraTS 2017 proposed a scheme that ensembles DeepMedic, FCN, and UNet models to minimize the bias introduced by using each single model and to improve the segmentation robustness \cite{kamnitsas2017ensembles}. For the second place in BraTS 2017, \cite{wang2017automatic} adopted a cascaded training scheme with 3 similar CNNs, each segmenting one of three tumour sub-regions. Each subsequent network takes the cropped output from the previous network as the input. The subsequent network is trained to segment a different tumour sub-region from the input. Most well performing submissions also adopted the ensemble learning method to minimize the bias introduced by different network architectures, hyper-parameters and loss functions.

\subsubsection{Proposed network architectures}
\begin{figure}[ht]
\includegraphics[width=\textwidth]{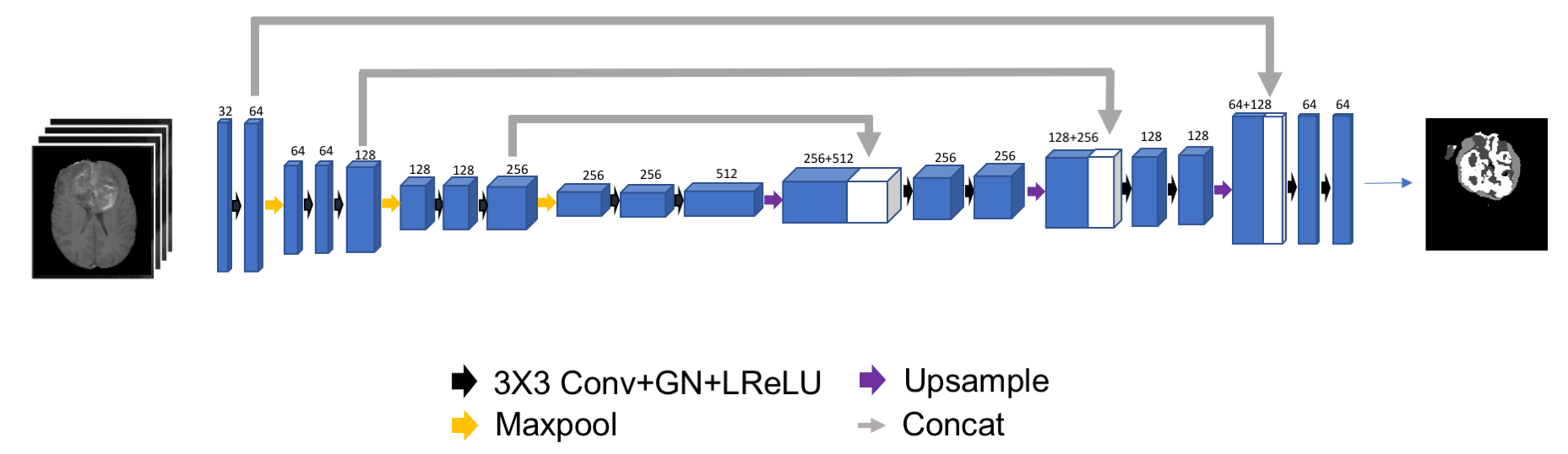}
\caption{3D UNet for tumour segmentation.} \label{unet}
\end{figure}

Given the good performance of UNet in the previous BraTS challenges, we chose the vanilla UNet as one of the architecture used in this work, shown in Figure~\ref{unet}. Unlike many previous submissions that used transposed convolution block in the decoder, we use linear upsampling block to reduce the number of parameters and save GPU memory for training. For activation function we empirically chose group normalisation.

\begin{figure}[ht]
\includegraphics[width=\textwidth]{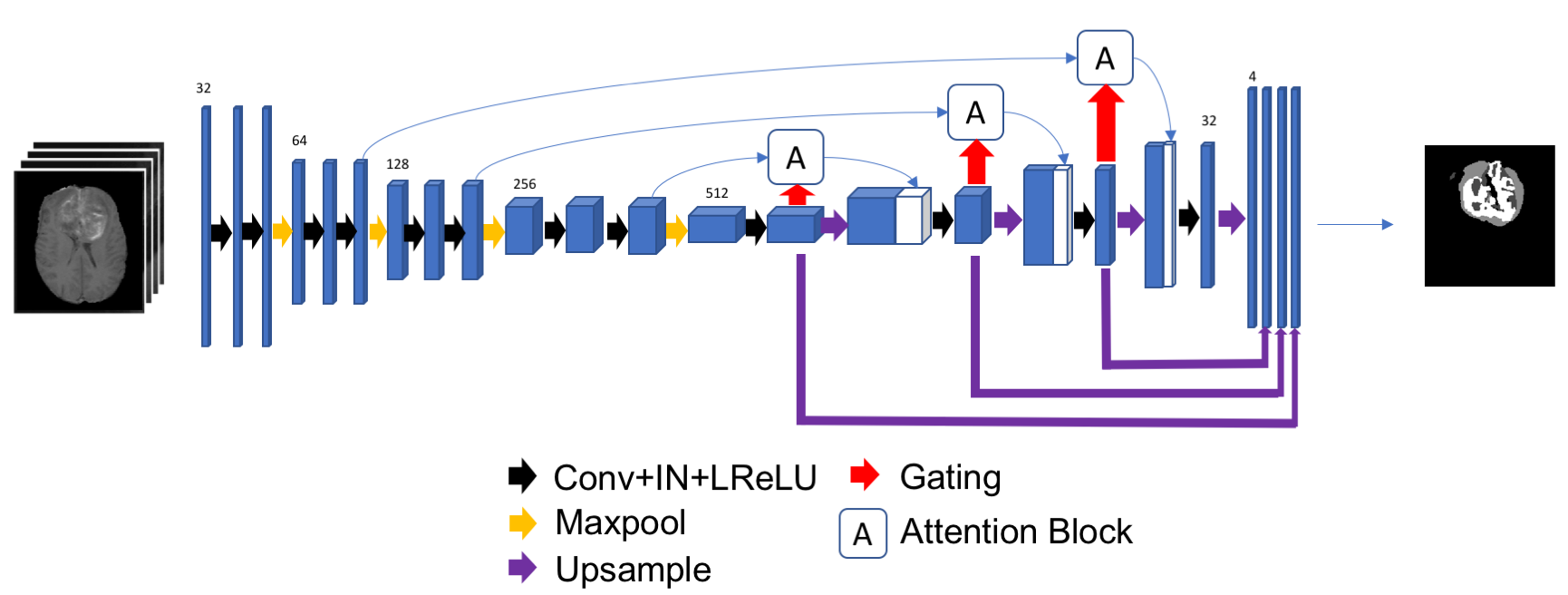}
\caption{3D UNet with attention gates for tumour segmentation.} \label{unet_att}
\end{figure}

The second architecture we used is a UNet with attention blocks (UNet-AT), shown in Figure~\ref{unet_att}. The attention mechanism has been demonstrated to be effective in improving the network performance across different tasks \cite{wang2017residual,wu2019u}. Our attention UNet model leverages the ability of attention blocks to concentrate on components that are more informative to achieve a better segmentation performance. Instance normalisation was empirically chosen for the attention UNet. Both UNet and UNet-AT used leaky ReLU with a leakiness of 0.01 as the activation function.

\subsubsection{Image pre-processing and augmentation}
It was pointed out in \cite{kamnitsas2017ensembles} that different image normalisation methods have not shown significant impact on segmentation performance. Therefore we simply applied z-score normalisation onto the raw images. Random image rotation and horizontal flipping were performed for data augmentation during training.

\subsubsection{Ensemble of network models}
We trained a number of different models (Table~\ref{ensemble}) for building the ensemble, using different loss functions (cross entropy or soft Dice) and post-processing steps (with or without conditional random field (CRF)). Due to the constraint of GPU memory, it was not possible for us to use the whole 3D volume as the input to any of our networks, so patch extraction was used for creating the training samples. The patches were randomly extracted with a 50\% chance from background and the other 50\% chance from any of the tumour sub-regions. Adam optimizer was used for training the models with the learning rate set to $10^{-4}$ and weight decay set to $10^{-5}$. All the models were trained for 60,000 iterations and it took approximately 40 hours to train a model on Nvidia Titan X.

\subsubsection{Post-processing}
We empirically adopted some automatic post-processing methods to further improve accuracy of the prediction from ensemble of the models, including removing small isolated whole/enhancing tumour from the prediction, adjusting the size of tumour core in line with the size of enhancing tumour and filtering based on the intensity distributions of different tumour labels. The post-processing significantly improved the Dice score and Hausdorff distance of enhancing tumour and tumour core.

\begin{table}[]
\centering
\caption{List of models for ensemble.}\label{ensemble}
\begin{tabular}{@{\extracolsep{4pt}}cccccc}
\hline
Model & \begin{tabular}[c]{@{}c@{}}Network\\ structure\end{tabular} & \begin{tabular}[c]{@{}c@{}}Patch\\ size\end{tabular} & \begin{tabular}[c]{@{}c@{}}Batch\\  size\end{tabular} & \begin{tabular}[c]{@{}c@{}}Loss \\ function\end{tabular} & CRF   \\ \hline
1     & UNet                                                        & 96                                                   & 4                                                     & Cross Entropy                                            & Yes  \\ 
2     & UNet                                                        & 96                                                   & 4                                                     & Cross Entropy                                            & No \\ 
3     & UNet                                                        & 96                                                   & 4                                                     & Soft Dice                                                & Yes  \\ 
4     & UNet                                                        & 96                                                   & 4                                                     & Soft Dice                                                & No \\ 
5     & UNet-AT                                                     & 128                                                  & 2                                                     & Soft Dice                                                & No \\ 
6     & UNet-AT                                                     & 128                                                  & 2                                                     & Cross Entropy                                            & Yes  \\ \hline
\end{tabular}
\end{table}

\subsection{Image feature extraction}
\begin{figure}
\includegraphics[width=\textwidth]{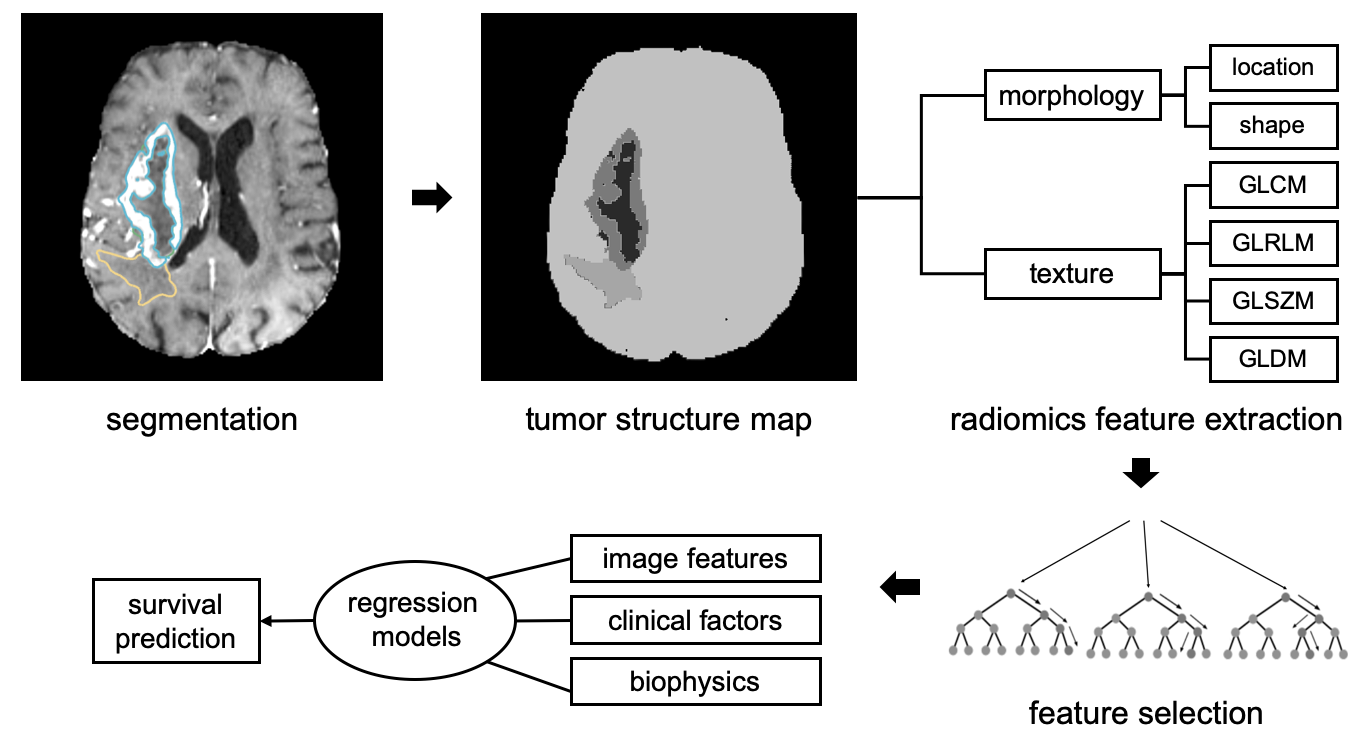}
\caption{Workflow of feature extraction and survival prediction.} \label{fig1}
\end{figure}

Based on the segmentation, we constructed a tumour structure map with four discrete values for each patient. The spatial distribution of tumour sub-regions provides the information of tumour heterogeneity and tumour invasiveness \cite{li2019decoding}. Quantitative features were extracted from the tumour structure map for survival prediction (Figure~\ref{fig1}). 

81 radiomics features were extracted for each region-of-interest (ROI), consisting of 13 morphological features and 68 texture features. The morphological features describe the location and shape of the ROI, including X-, Y-, Z-coordinates of ROI centroid with respect to the brain centroid, volume, surface area, surface-area-to-volume ratio, sphericity, maximal 3D diameter, major axis length, minor axis length, least axis length, elongation and flatness. The textures features reveal the spatial distribution of tumour sub-regions within the ROI, which include 22 grey level occurrence matrix (GLCM) featuress, 16 gray level run length matrix (GLRLM) features, 16 grey level size zone matrix (GLSZM) features and 14 gray level dependence (GLDM) features. We extracted features for two ROIs, the whole tumour (WT) and the tumour core (TC), which amounted to 162 features in total. Feature extraction was implemented with Python package \textit{PyRadiomics} \cite{van2017computational}.

\begin{figure}
\includegraphics[width=\textwidth]{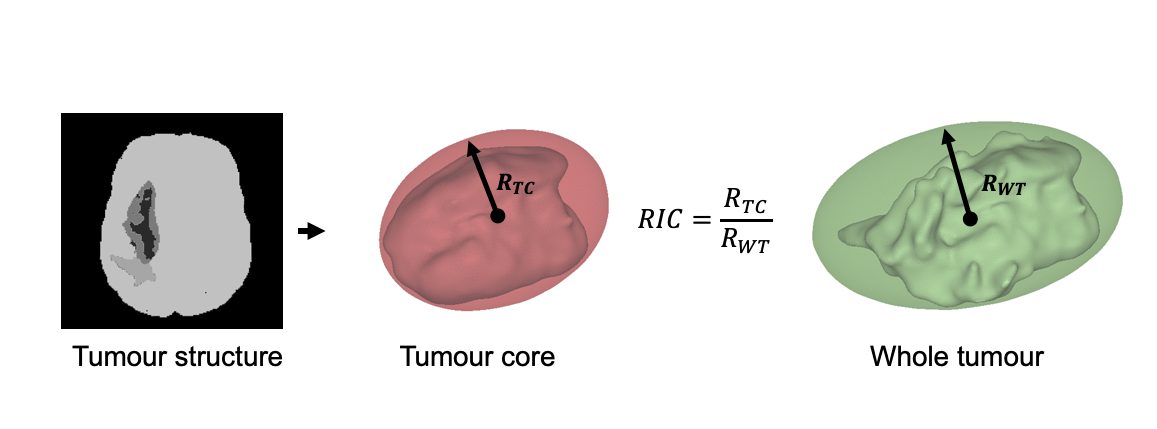}
\caption{Illustration of the calculation of relative invasiveness coefficient (RIC).} \label{figS3}
\end{figure}

Apart from radiomics features, we also considered the biophysics modelling of tumour growth \cite{baldock2014patient}. The relative invasiveness coefficient (RIC) is of particular interest, which is defined as the extent ratio between the hypoxic tumour core and infiltration front according to the profile of tumour diffusion \cite{li2019characterizing}. The characteristic extent of each ROI is calculated from the minimum volume ellipsoid. In this study, the ratio of the second semi-axis length between TC and WT was calculated as the RIC (Figure~\ref{figS3}).

\subsection{Feature selection}
The large number of extracted radiomics features provide the capability for prognostic modelling, while it poses a risk of over-fitting in such a small training set. We apply feature selection techniques on the training set seeking a subset of radiomics features.

First, we detected highly correlated features (Pearson r$>$0.95) and removed redundant features. The number of features was further reduced through a recursive feature elimination (RFE) scheme with a random forest regressor. The feature importance was evaluated and less important features were eliminated iteratively. The optimal number of features was determined by achieving the best performance on cross-validation results. The feature selection procedure was performed with the R package \textit{caret} \cite{JSSv028i05}.

\subsection{Prognostic models}
We compared three prognostic models, the baseline model which only uses age, the radiomics model and the biophysics-driven tumour invasiveness model. \\ \\
\noindent \textbf{Baseline model}
Age is the only available clinical factor and significantly correlated with the survival time (Pearson r=-0.486, p$<$1e-5) on the training set. We constructed a linear regression model using age as the only predictor. \\
 
\noindent \textbf{Radiomics model}
The selected radiomics features and age were integrated into a random forest (RF) model. This data-driven model included the largest number of image features.
\\

\noindent \textbf{Tumour invasiveness model}
RIC derived from the tumour structure map was used to describe tumour invasiveness. An epsilon-support vector regression ($\epsilon$-SVR) model was built using age and RIC as two predictors.  \\

Classification and regression metrics were used to evaluate the prognostic performance. For classification, overall survival time was quantitised into three survival categories: short survival ($<$10 months), intermediate survival (10-15 months) and long survival ($>$15 months). The 3-class accuracy metric wass evaluated. Regression metrics include the mean squared error (MSE), median squared error (mSE) and Spearman correlation coefficient $\rho$. 

\section{Results}
\subsection{Segmentation performance}
The segmentation performance on BraTS 2019 validation dataset is reported in Table~\ref{segres}, in terms of Dice score and Hausdorff distance. Among each individual models, model 1 (UNet with cross entropy loss and CRF) achieves the highest Dice scores of whole tumour and enhanced tumour. For Dice score of tumour core, model 2 (UNet with soft Dice loss and CRF) gives the best result. The ensemble of all six models gives the best result across all metrics. The ensemble of models significantly improves the overall performance of the models. Although the performance of UNet-AT (models 5, 6) is not as good as UNet (models 1-4), but adding UNet-AT improves the overall performance of the ensemble. 

\begin{figure}
\includegraphics[width=\textwidth]{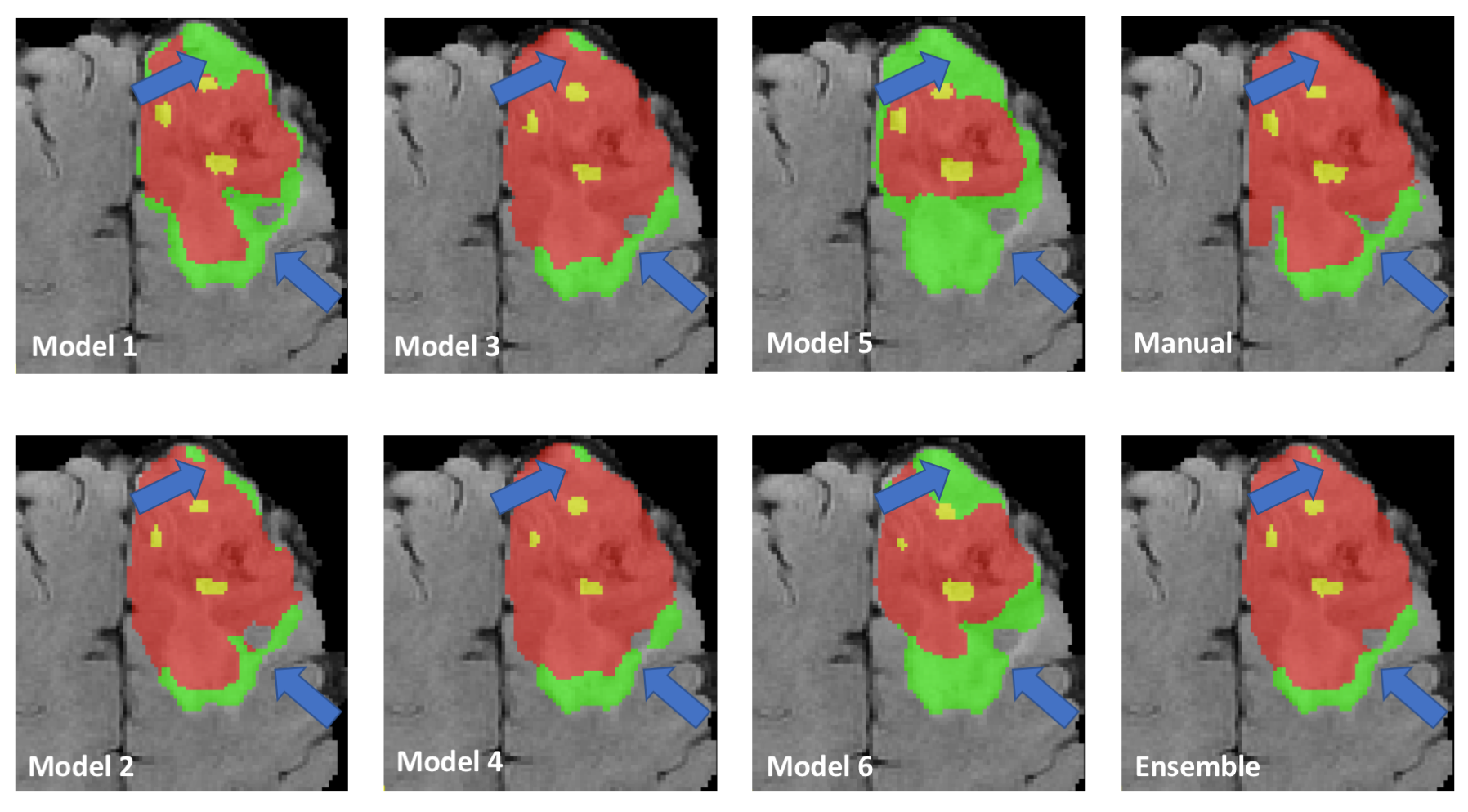}
\caption{Segmentation result of the brain tumour from a training image in BraTS 2019 segmented by 6 models and the ensemble model, where green depicts oedema, red depicts tumour core, and yellow depicts enhancing tumour.} \label{labels}
\end{figure}

An example showing improvement of using ensemble of models is given in Figure \ref{labels}

The performance of the ensemble of models on test dataset is given in Table~\ref{segrestest}. 

\begin{table}
\caption{Segmentation result on validation set.}\label{segrestest}
\begin{tabular}{ccccccc}
\hline
Model   & Dice\_ET                         & Dice\_WT                         & Dice\_TC                         & Hausdorff\_ET & Hausdorff\_WT & Hausdorff\_TC \\ \hline
1        & 0.74                           & 0.90                           & 0.78                           & 6.56         & 5.84         & 8.66         \\ 
2        & 0.71                           & 0.89                           & 0.79                           & 6.08         & 5.37         & 8.19         \\ 
3        & 0.73                           & 0.89                             & 0.80                           & 6.68         & 6.93         & 7.50         \\ 
4        & 0.73                           & 0.89                           & 0.80                           & 7.06         & 7.05         & 7.89         \\ 
5        & 0.70                            & 0.89                           & 0.73                           & 6.14         & 8.61         & 13.13        \\ 
6        & 0.72                           & 0.89                           & 0.77                           & 6.00         & 4.81         & 7.99         \\ 
Ensemble & 0.75 & 0.90 & 0.81 & 4.99         & \textbf{4.70}         & 7.11         \\ \hline
Post- \\processing & \textbf{0.79} & \textbf{0.90} & \textbf{0.83} & \textbf{3.37}         & 5.04         & \textbf{5.56}         \\ \hline
\end{tabular}
\end{table}

\begin{table}
\caption{Segmentation result on test set.}\label{segres}
\begin{tabular}{ccccccc}
\hline
Label  & Dice\_ET & Dice\_WT & Dice\_TC & Hausdorff\_ET & Hausdorff\_WT & Hausdorff\_TC \\ \hline
Mean   & 0.82  & 0.88  & 0.82  & 2.55         & 5.49         & 4.80         \\
StdDev & 0.18  & 0.12  & 0.26  & 4.53         & 7.19         & 8.38         \\
Median & 0.85  & 0.92  & 0.92  & 1.73         & 3.16         & 2.24         \\ \hline
\end{tabular}
\end{table}

\subsection{Survival prediction}
\subsubsection{Subset of radiomics features}
One morphological feature and four texture features were selected through RFE. The selected radiomics features were ranked according to feature importance (Table \ref{tab1}), which were included into the radiomics prognostic model. \\

\begin{table}[]
\centering
\caption{Selected radiomics features.}\label{tab1}
\begin{tabular}{ccllc}
\hline
Rank\ \ & Region\ \ & Categories \ \ & Feature name \ \          & Importance (\%) \\ \hline
1       & TC        & Texture        & glcm\_ClusterShade        & 100             \\
2       & WT        & Texture        & glcm\_MaximumProbability  & 87              \\
3       & TC        & Texture        & glcm\_SumSquares          & 79              \\
4       & TC        & Texture        & glszm\_MaximumProbability & 70              \\
5       & TC        & Morphology     & shape\_center\_Z          & 55              \\
\hline
\end{tabular}
\end{table}

\subsubsection{Prognostic performance} 
The training set includes 101 patients with Gross Tumour Resection (GTR) and the validation set includes 29 patients. We first trained the models on the full training set and evaluated on the training set. This usually overestimates the real performance so we also repeated ten-fold cross-validation to assess the generalisation performance on unseen data. The training set was split into ten folds where models were trained on nine folds and evaluated on the hold-out fold. The training performance and cross-validation performance are reported in Table~\ref{tab2}. Finally, we evaluated the performance of the trained models on the independent validation set (Table~\ref{tab3}).

\begin{table}[]
\centering
\caption{Prognostic model performance on the training set.} \label{tab2}
\begin{minipage}{\textwidth}
\begin{tabular}{@{\extracolsep{4pt}}lcccccccc}
\hline
\multirow{2}{*}{Model} & \multicolumn{4}{c}{Training}  & \multicolumn{4}{c}{Cross-validation}  \\ \cline{2-5} \cline{6-9}
 & Accuracy & MSE  & mSE & $\rho$ & Accuracy & MSE & mSE & $\rho$ \\ \hline
Baseline & 0.48 & 88822 & 21135 & 0.48 & 0.48 & \textbf{93203} & 21923 & 0.47   \\ 
Radiomics & \textbf{0.73} & \textbf{22249} & \textbf{6174}  & \textbf{0.93}  & 0.47 & 103896 & 31265 & 0.37 \\
Invasiveness & 0.51 & 95728 & 17165 & 0.49  & \textbf{0.50} & 99707 & \textbf{18218}  & \textbf{0.47}  \\ 
\hline

\end{tabular}
\end{minipage}
\end{table}

\begin{figure}
\includegraphics[width=\textwidth]{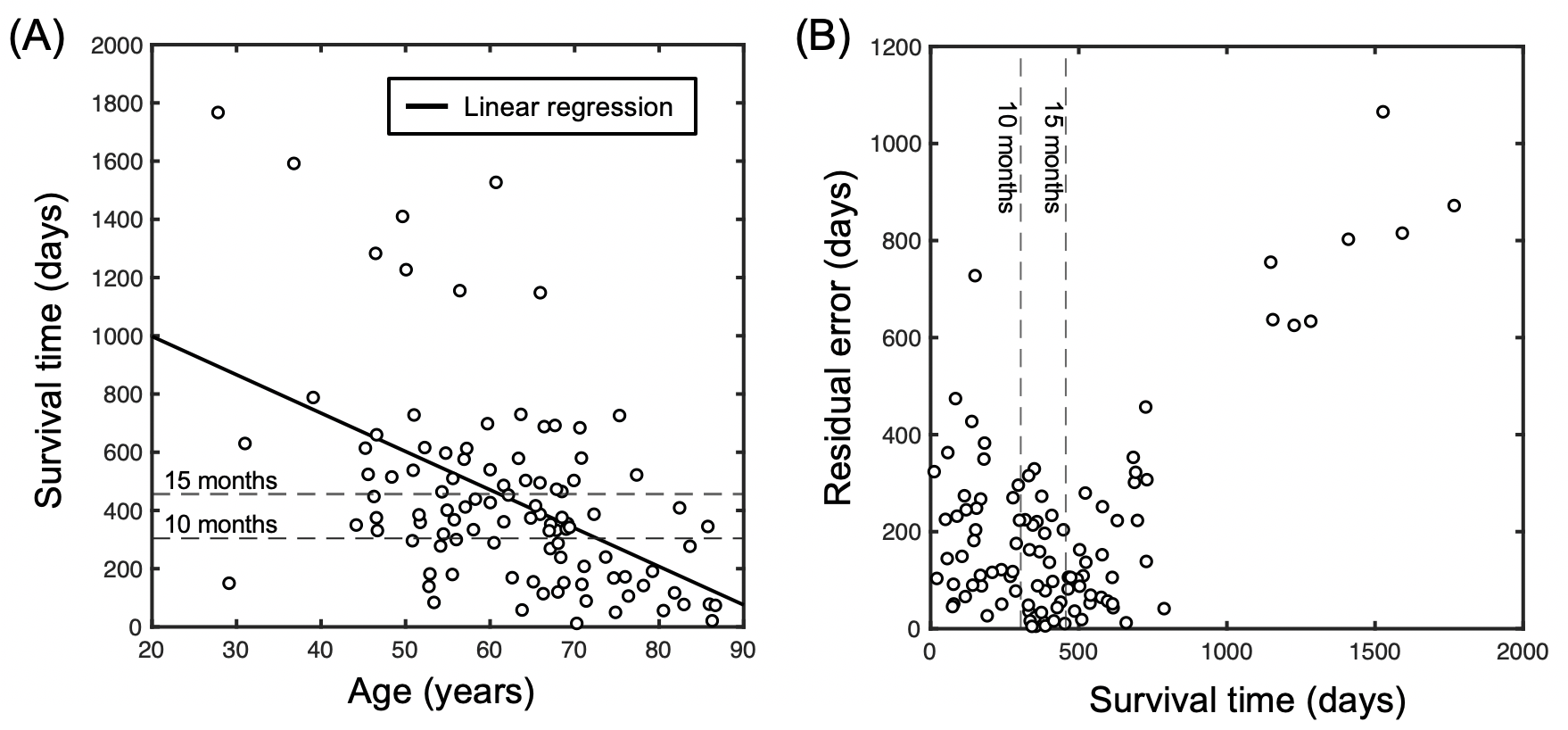}
\caption{Age-only linear regression results (A) and the distribution of residual error on the training set (B). The circle dots and straight line are the training data and prediction results respectively. Two dash lines represents the two thresholds for short-, mid- and long-survival categories.} \label{fig2}
\end{figure}

The radiomics model performed the best when tested on the same training data, reaching an accuracy of 0.73. However, the performance dropped significantly on the cross-validation results and independent validation set, highlighting the over-fitting problem. In contrast, the baseline model and invasiveness model show good capabilities of generalisation on unseen datasets. The linear fitting result of the baseline model and prediction error are shown in Figure \ref{fig2}. Large errors are found for patients survived more than 1,000 days. The invasiveness model outperformed the baseline model with an accuracy of 0.50 on the cross-validation and 0.59 on the independent validation set. We chose the invasiveness model to submit for test set evaluation. The model won  the  second  place  of  the  MICCAI  2019  BraTS  Challenge  for  the overall survival prediction task with an accuracy of 0.56. The leader board is available at \url{https://www.med.upenn.edu/cbica/brats2019/rankings.html}.

\begin{table}[]
\caption{Prognostic model performance on the independent validation set.} \label{tab3}
\begin{minipage}{\textwidth}
\centering
\begin{tabular}{@{\extracolsep{4pt}}lcccc}
\hline
\multirow{2}{*}{Model} & \multicolumn{4}{c}{Validation set}  \\ \cline{2-5}
& Accuracy & MSE  & mSE & $\rho$ \\ \hline
Baseline & 0.45 & 90109 & 36453  & 0.27  \\
Radiomics & 0.48 & 97883 & \textbf{29535} & 0.28   \\
\textbf{Invasiness}$^{\ast}$ & \textbf{0.59} & \textbf{89724} & \textbf{36121}  & \textbf{0.36}  \\ \hline
\multicolumn{5}{l}{$^{\ast}$This model is submitted for evaluation on test set.} 
\end{tabular}
\end{minipage}
\end{table}

\section{Discussion}
For tumour segmentation task, \cite{isensee2018no} demonstrated that the generic UNet with a set of carefully selected hyperparameters and a pre-processing method achieved the state-of-the-art performance on brain tumour segmentation. Our experiments results further confirm this. However, \cite{isensee2018no} adopted co-training scheme to further improve the UNet performance on segmenting enhancing tumour and our models also encountered some difficulties when segmenting enhancing tumour for some subjects. Including networks with different or more complex structure in our ensembles has been demonstrated to be effective to alleviate this issue to some extent.

The performance of our method on test dataset did not get into top three in the end. Judging from the Dice score and Hausdorff distance we achieved, we suspect including 6 models in the ensemble may not necessary and may introduce a higher false negative rate in the prediction of whole tumour.

For survival prediction, our invasiveness model achieved the best prognostic performance using age and RIC as predictors. Although more complex features were integrated in the radiomics model, it did not outperform the invasiveness model or even the baseline model on validation sets. This highlights the over-fitting risk of data-driven model and demonstrates the advantage of features from biophysics modelling. The RIC feature is designed from the diffusion equation of tumour growth, reflecting the physiological information of tumour infiltration. We note that RIC can be calculated from the 161 radiomics features but relevant features were not selected in the data-driven approach. This suggested that a better feature selection algorithm may be needed. However, it is a challenging task to select radiomics features due to several difficulties. First, the 'large n small p' problem is underdetermined and poses a risk of taking none-sense features by chance. On the other hand, high-order radiomics features are usually sensitive to the intensity distribution and image noise, limiting the performance on unseen data set. The reproducibility and robustness of radiomics feature should be examined in future studies. Moreover, the radiomics features are difficult to explain, which prevents the application in clinical practice. The biophysics modelling and other prior knowledge could be integrated in the feature design and selection schemes.

In this study, we assume that the boundary of WT represents the tumour infiltration front while the boundary of TC represents the active proliferation. Prognostic value of this image-derived feature was verified in the regression of survival time. However, the four structural modalities of MRI (e.g. T1-Gd, T1, T2 and T2-FLAIR) are not capable to reflect the physiological process. Advanced MRI modalities such as perfusion weighted imaging (PWI) and diffusion tensor imaging (DTI) could be used for better assessment of tumour heterogeneity and invasiveness \cite{li2018intratumoral}.

It is also noted that all the three prognostic models achieved lower mSE than MSE, which indicates the skewed distribution of prediction error. The uni-variate linear regression fits well for the mid-survival patients, while large errors were observed for short- and long-survival patients (Fig \ref{fig2}). Weighted loss or appropriate data transformation could be used to reduce the influence of long-tail survival distribution in future studies.

\section{Conclusion}
We have developed a deep learning framework for automatic brain tumour segmentation and a biophysics-guided prognostic model that performs well for overall survival prediction of patients.

\section{Acknowledgement}
This work was supported by the SmartHeart EPSRC Programme Grant (EP/P001009/1) and  the NIHR Imperial Biomedical Research Centre (BRC) .

\bibliographystyle{splncs04}
\bibliography{ref}

\end{document}